\documentclass{blois}

\def\be{\begin{equation}}
\def\ee{\end{equation}}
\def\bea{\begin{eqnarray}}
\def\eea{\end{eqnarray}}

 \newcommand{\Photo}{}

\begin{document}
\vspace*{4cm}
\title{The Pierre Auger Observatory: new results and prospects}

\author{Lorenzo Cazon$^1$,for the Pierre Auger Collaboration$^2$}

\address{$^1$ LIP, Av. Prof. Gama Pinto 2, Lisbon, Portugal,   \\  $^2$ Observatorio Pierre Auger, Av. San Mart\' \i n Norte 304, 5613 Malarg\" ue, Argentina \\ E-mail$^1$: cazon@lip.pt \\E-mail$^2$: auger$\_$spokespersons@fnal.gov \\Full author list: http://www.auger.org/archive/authors$\_$2017$\_$10.html}

\maketitle\abstracts{ Ultra High Energy Cosmic Rays are the most energetic particles directly measured, reaching orders of magnitude above those attained in the LHC. The Pierre Auger Observatory is shedding light onto the long-standing mysteries of their nature and origin. The present paper briefly reviews the latest results: the energy spectrum and mass composition of  UHECR, limits on the fluxes of photons and neutrinos,  the study of anisotropies in the arrival directions of UHECS, and finally the constraints on high energy hadronic models. The paper will finish with the upgrade of the Observaotry: AugerPrime.}

\section{Introduction}

Ultra High Energy Cosmic Rays (UHECR) reach up to 400 times the energy of protons in the LHC at the center of mass, providing a window of opportunity to study the high energy frontier of the cosmos. The Pierre Auger Observatory, which is making great leaps in the long standing mystery of UHECR nature and origin, is the largest and most complete experiment in the field. It is located in the Argentinian region of Pampa Amarilla, a high plateau at 1400 m.a.s.l. right at the East side of the Andes. It covers an area of 3000 km$^2$ with 1660 water-Cherenkov stations (called, the surface detector or SD), and it is overlooked by 24 fluorescence detectors (called FD).

When a UHECR hits an atom at the top of the atmosphere, it triggers a cascading reaction of particles, which is called an Extensive Air Shower (EAS).  EAS charged particles excite the nitrogen in their passage through the atmosphere, emitting fluorescence light (detected by the FD), and eventually they trigger the water-Cherenkov detectors at the ground surface.  FD allows an almost calorimetric measurement of the energy of the primaries, but it only operates in clear, moonless nights, resulting in a duty cycle of $\sim$ 13\%. On the other hand, the SD has a duty cycle of almost 100\%, providing high statistics. The hybrid nature of the observatory allows the energy calibration of SD with the sub-sample of events coming from FD, combining high statistics with high accuracy.

\section{Energy Spectrum and mass composition}

The latest results show  that the break in the energy spectrum commonly called the {\it ankle} is  observed at  $E_{\rm ankle}=5.08\pm0.06\pm0.8(sys)$ EeV. Below this energy, the spectrum is well described by a power law $E^{-\gamma}$ with $\gamma=3.293 \pm0.002 \pm 0.05(sys)$, changing by $\Delta \gamma=-0.76$ after the {\it ankle}. There is then a strong suppression of the flux.  The energy at which the integral flux drops a factor of 2 with respect to no suppression, is located at $E_{1/2}=23\pm 1 \pm 4(syst)$ EeV \cite{RefSpectrum0}. If the primaries were extragalactic protons, the GZK effect in which they would undergo energy losses though photo-pion production, would result in $E_{1/2}=53$ EeV, a value at odds with observations. 

\begin{figure}
  \begin{minipage}{0.5\linewidth}
    \includegraphics[width=3in]{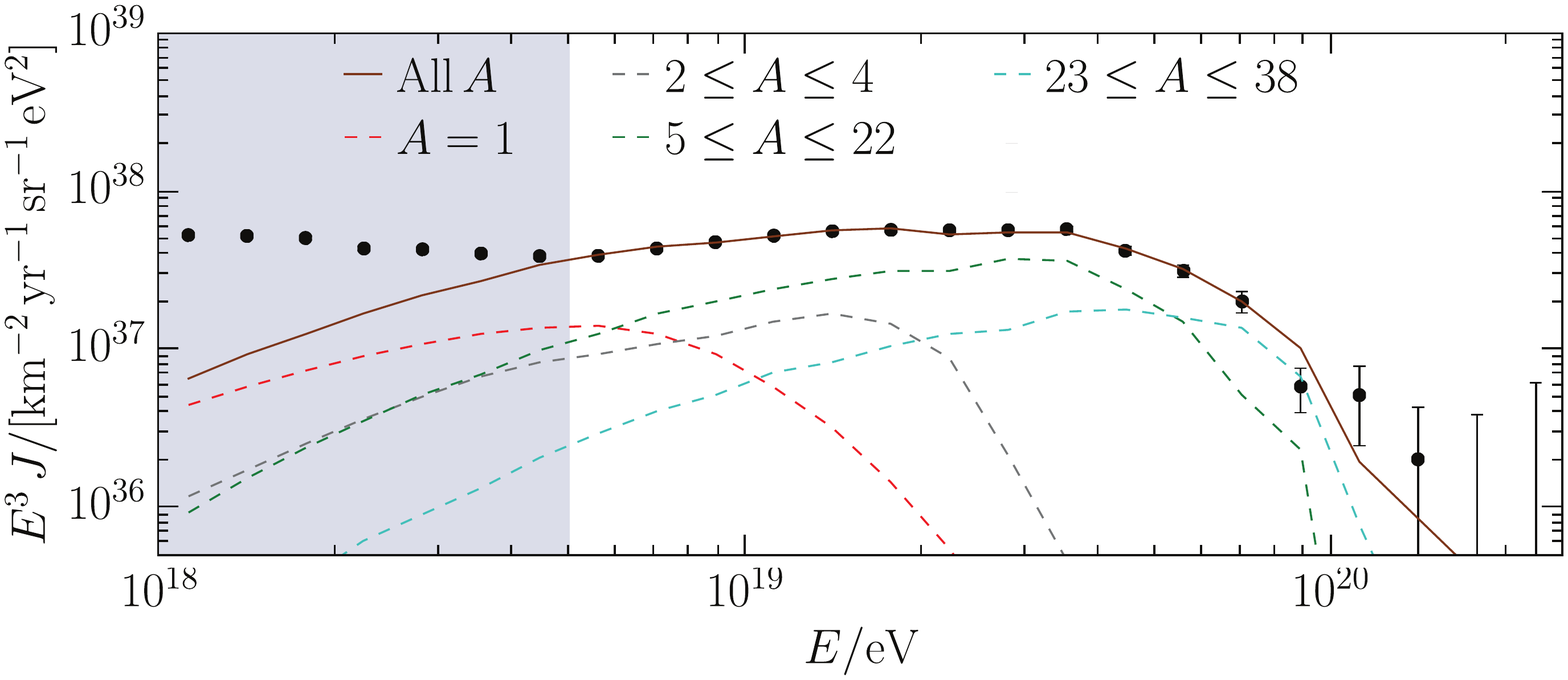}
      \end{minipage}
  \begin{minipage}{0.5\linewidth}
    \includegraphics[width=3in]{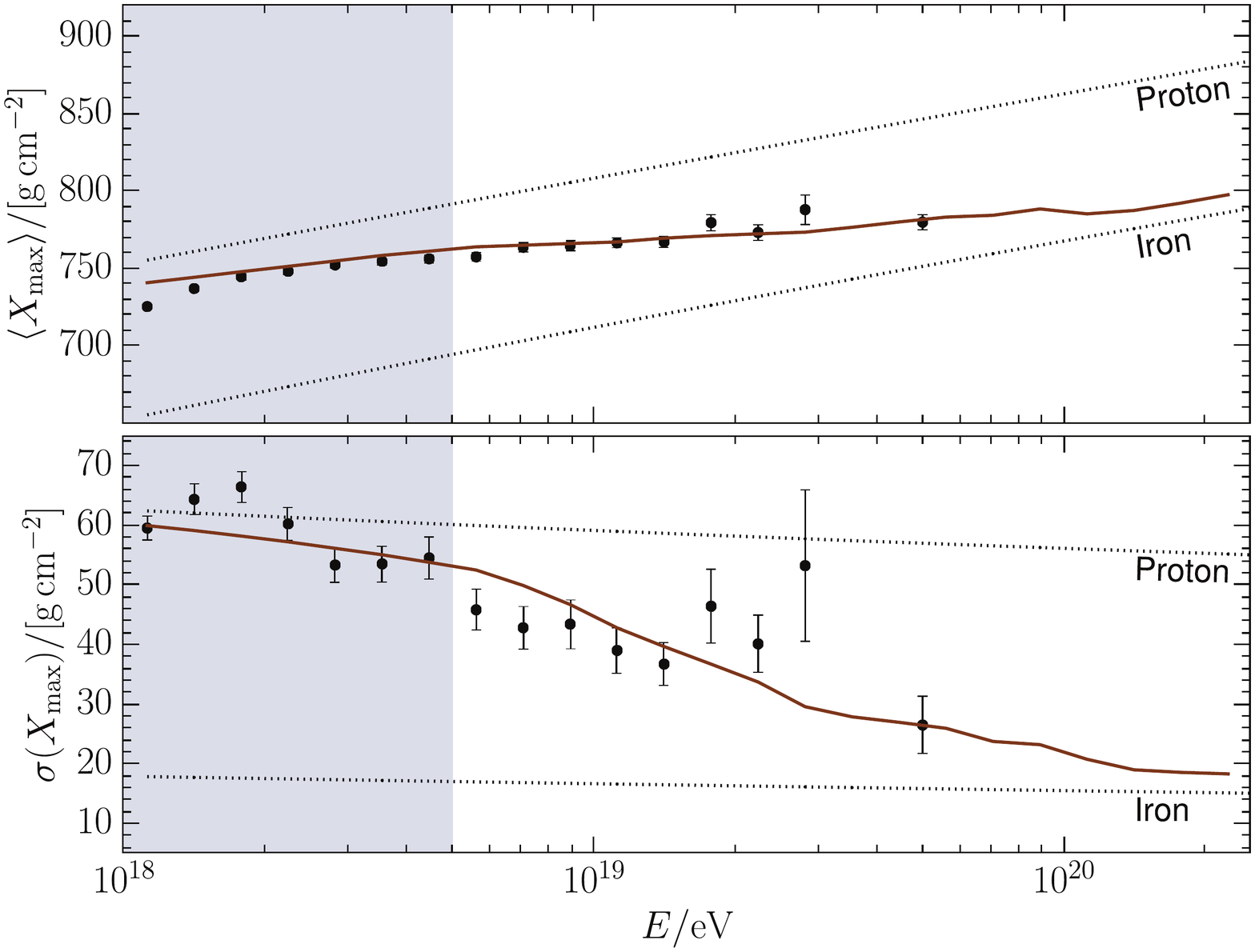} 
  \end{minipage}
  \caption[]{Energy spectrum (left) and $\langle X_{max} \rangle$ and $\sigma(X_{\rm max})$ (right) as a function of the energy of the primary. Continuous line represent a global fit performed in the energy range above the shaded region, which includes the effect of propagation and magnetic fields starting from an injection spectrum and composition for a discrete distribution of sources. \cite{RefSpectrum1,RefSpectrum2}}
  \label{SpectrumMassFit}
\end{figure}

The best parameter to access the composition of the UHECR is  $X_{\rm max}$, the depth at which the number of charged particles reaches a maximum. The analysis of the average number $\langle X_{\rm max} \rangle$ and its fluctuations $\sigma(X_{\rm max})$ as a function of the energy are shown in Figure \ref{SpectrumMassFit}, right panel. Light primaries penetrate deeper in the atmosphere, and also fluctuate more than heavier primaries.
It can be concluded that the average mass of cosmic rays evolves towards a lighter composition between $10^{17.2}$ and $10^{18.33}$ eV. At higher energies the trend is inverted and the average mass increases with energy \cite{RefMass}.


Based on elaborate 4D simulations of the propagation of UHECR we have studied  the energy spectrum  and chemical composition of the UHECR at their sources  that best fit the spectrum and chemical composition as measured by the Pierre Auger Observatory.
We found that the details of the local large scale structure are of minor importance for the derived parameters of the source spectra. On the other hand, by including the the diffusion caused by the EGMF one goes from $E^{-1}$ to $E^{-1.6}$ injection spectrum at the sources, caused by the suppression of cosmic ray flux at low rigitidites.
Finally, Pierre Auger
Observatory data suggest that the chemical composition of the UHECR at their sources is dominated
by intermediate-mass nuclei.


\begin{figure}
  \begin{minipage}{0.5\linewidth}
    \includegraphics[width=3in]{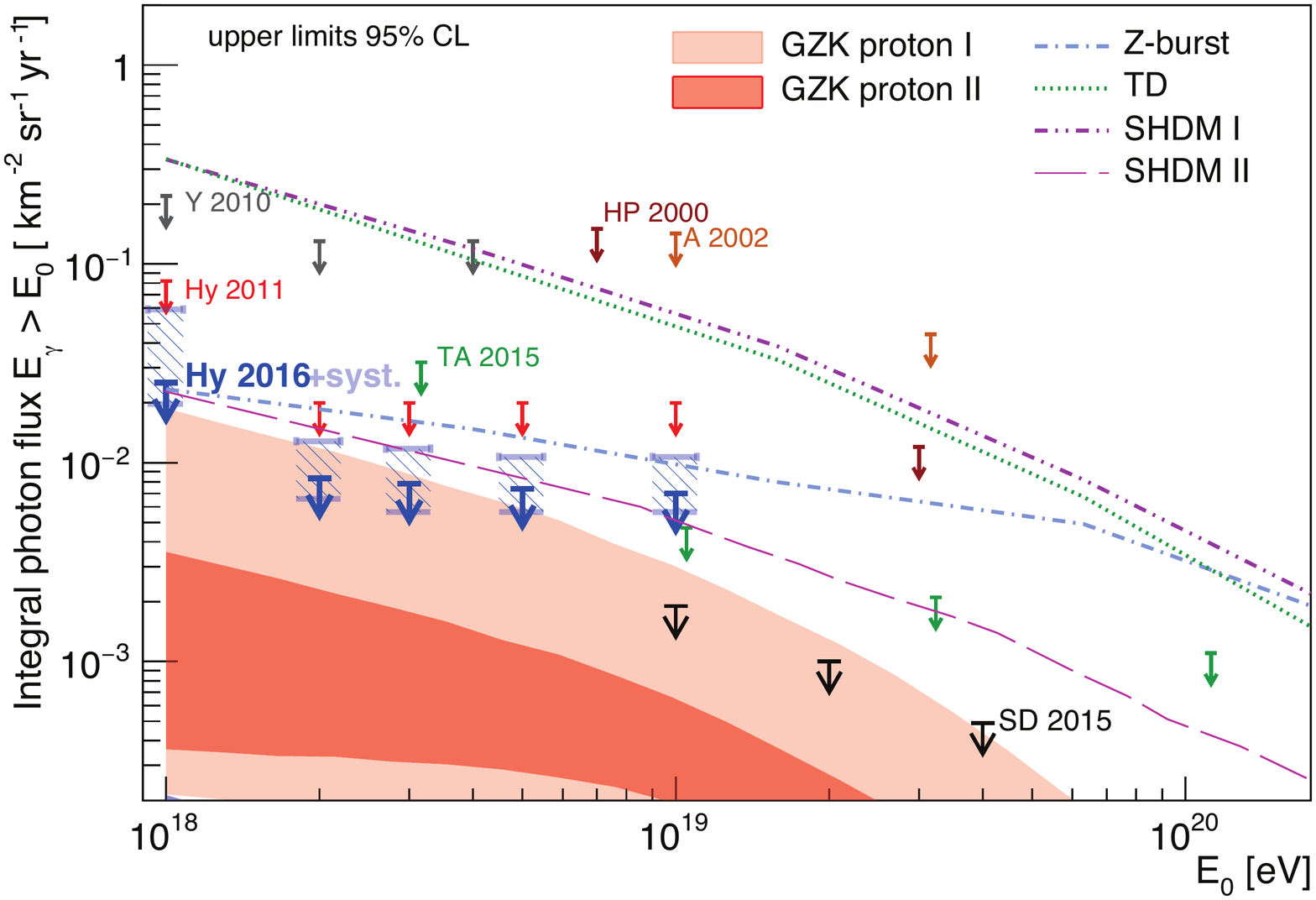}
  \end{minipage}
  \begin{minipage}{0.5\linewidth}
    \includegraphics[width=3in]{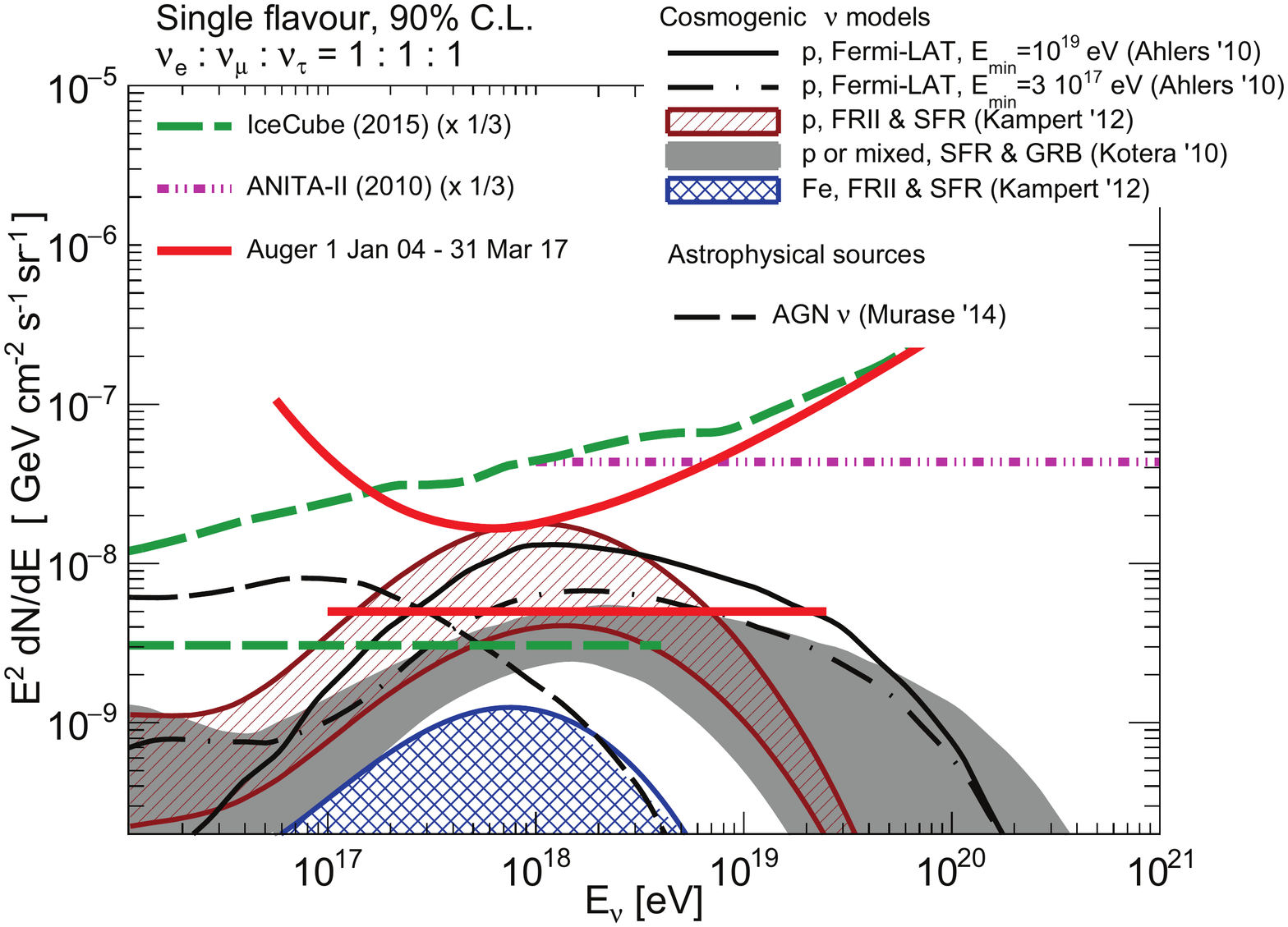}
  \end{minipage}
  \caption[]{Upper limit on the diffuse photon (left) and neutrino (right) flux compared to limits from other experiments and model predictions \cite{RefPhotons,RefNeutrinos} }
  \label{PhotonsNeutrinos}
  \end{figure}

The upper limits on the diffuse flux of photons and neutrinos constrain top-down models for the origin of UHECRs and are approaching the flux expected from the GZK effect, produced after the decay of a pion from the proton - CMB interaction, see Fig. \ref{PhotonsNeutrinos}.
A search for point photon and neutrino sources has also been performed over different sky targets, yielding no evidence for emitters in any of the studied source classes \cite{RefPN1,RefPN2,RefPN3}. 

\section{Anisotropies}

The Pierre Auger Collaboration recently announced the observation of a large-scale anisotropy in the arrival direction of cosmic rays above $8 \times 10^{18}$ eV \cite{RefScience}. A significance of 5.4 $\sigma$  was achieved in the amplitude of the first harmonic in right ascension. The events in a lower energy bin (4 $<$ E $<$8 EeV) were found to be compatible with isotropy. A sky-map of the intensity of cosmic rays arriving above 8 EeV is shown in Fig. \ref{MapHad}, left panel. The data can be described with a dipole with a total amplitude of $6.5^{+1.3}_{-0.9}$\%.  The direction of the dipole points towards $(l, b)=(233^\circ,-13^\circ$) and it s indicated with a star in Fig \ref{MapHad}, which is $125^\circ$ off the Galactic Center.
The  flux-weighted distribution of nearby galaxies as mapped by the 2MASS redsfhit survey (2MRSS)  is shown as an open diamond in Fig. \ref{MapHad}. Also shown is the effect of the galactic magnetic field using a particular model and two rigidites compatible with the composition fractions observed by Auger, showing improved agreement between the UHECR dipole and expectation if sources were a fair sample of 2MRS galaxies.


\section{Constraints to hadronic models}

\begin{figure}
  \begin{minipage}{0.5\linewidth}
    \includegraphics[width=3in]{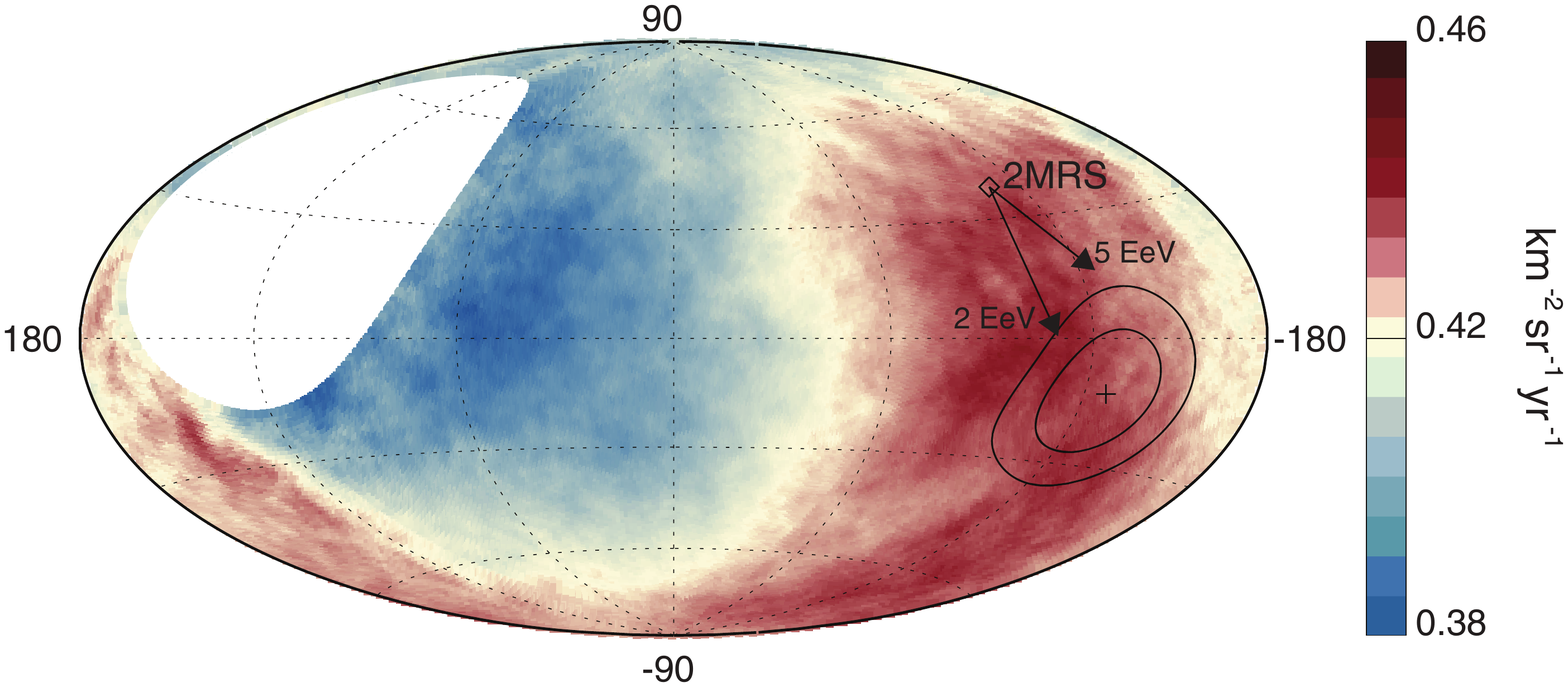}
  \end{minipage}
  \begin{minipage}{0.5\linewidth}
    \includegraphics[width=2.7in]{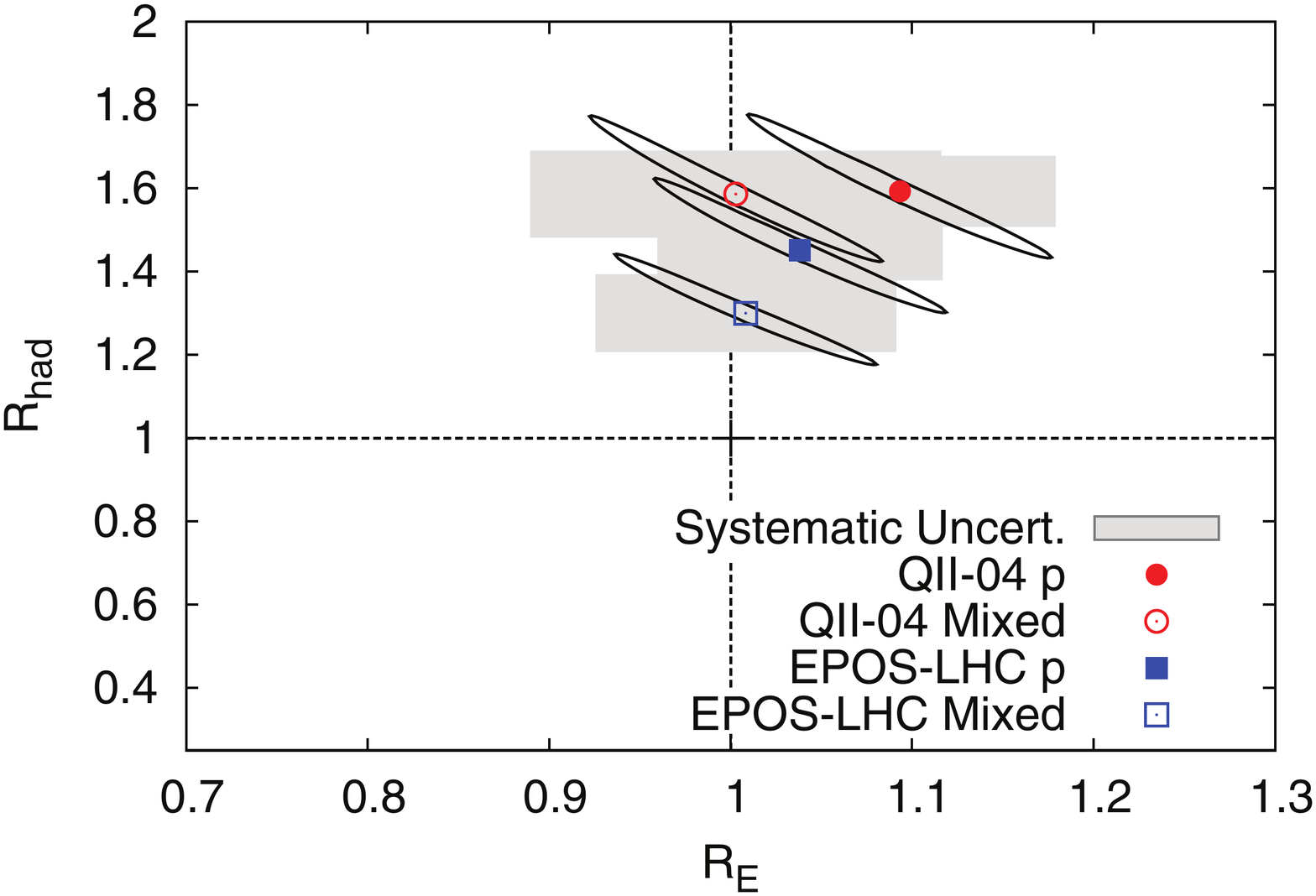}
  \end{minipage}
  \caption[]{Left: map of the intensity in galactic coordinates for E>8 EeV, smoothed by 45$^\circ$. The reconstructed dipole is indicated with a star, and the contours the 68\% and 98\% C.L. regions.  Also indicated is the dipole of the 2MRS catalogue, and arrows show indicate the dipole as modified by a Galactic magnetic field model, for two $E/Z$ rigidities compatible with Auger composition observations. Right panel: Best-fit values of electromagnetic and hadronic scaling for QGSJet-II-04 and EPOS-LHC, for pure proton (solid circle, square) and mixed composition (open circle, square). The ellipses and grey boxes show the 1$-\sigma$ statistical and systematic uncertainties.}
  \label{MapHad}
\end{figure}

An EAS is composed mainly of an hadronic cascade, mostly pions, which evolves very close to the shower axis, and by the electromagnetic cascade, which stems basically from the decay of neutral pions, resulting in a decoupling of both cascades after just a few generations \cite{RefCazon}. When charged pions reach low energies, they most likely decay into muons, which act as carriers of information about the hadronic cascade. 

\begin{figure}
  \begin{minipage}{0.5\linewidth}
    \includegraphics[width=2.7in]{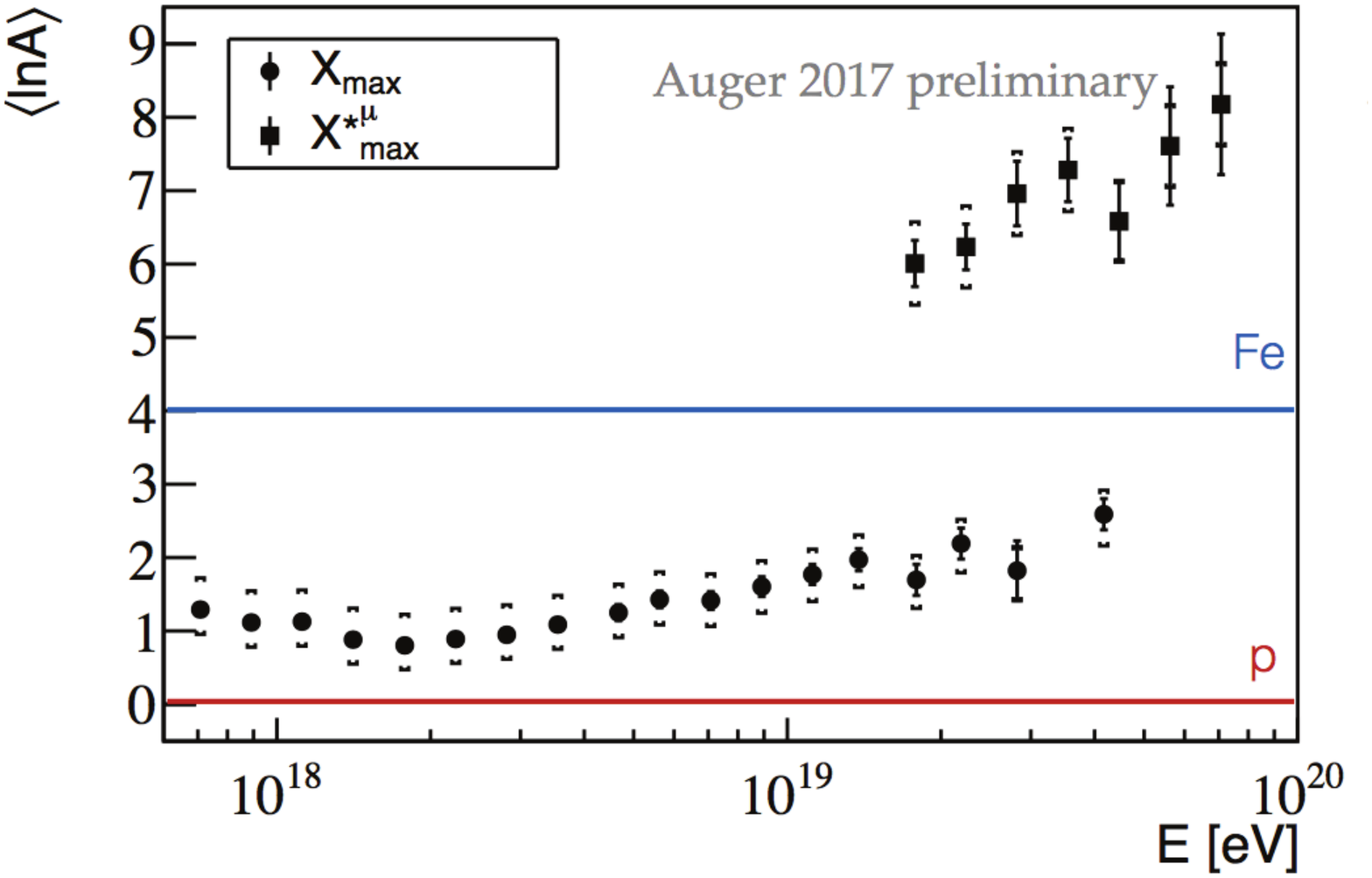}
  \end{minipage}
  \begin{minipage}{0.5\linewidth}
    \includegraphics[width=2.7in]{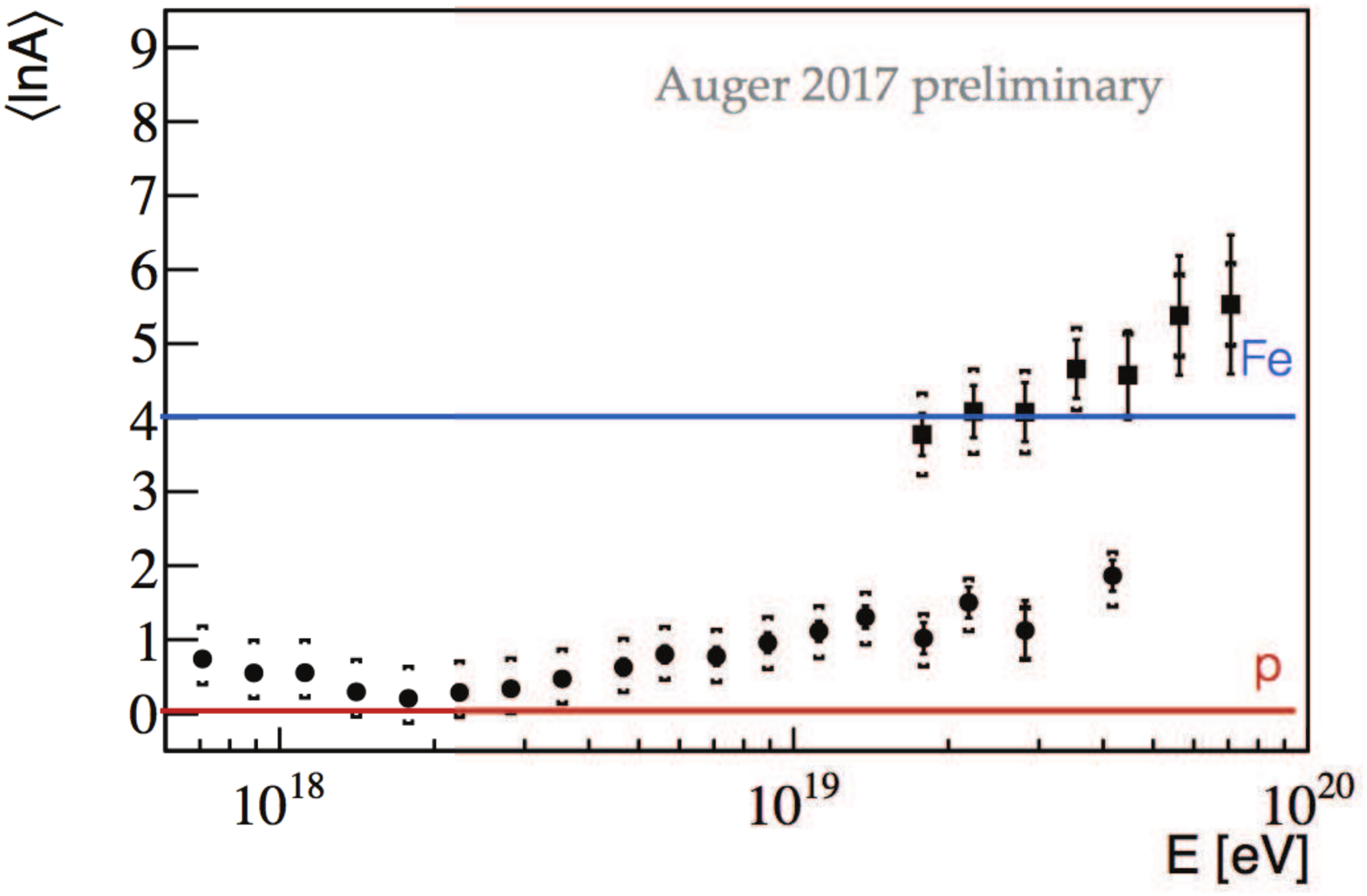}
  \end{minipage}
  \caption[]{The evolution with energy of $\langle ln A \rangle$ as obtained from the measured $X^\mu_{\rm max}$ (squares) and $X_{\rm max}$ (dots) . EPOS-LHC (left) and QGSJetII-04 (right) are used as reference models. Square brackets correspond to the systematic uncertainties.}
  \label{lnA}
 \end{figure}

The Pierre Auger Observatory has found that the number of muons in simulations is smaller than measured by a factor ranging between 1.3 to 1.7 \cite{RefHAS},\cite{RefPRL}, and also that those data do not favour a change in the EM energy scale by the FD that could account for such observation \cite{RefPRL}, see Fig. \ref{MapHad} right panel.  At the same time, it has been observed that models predict a depth where the muon production reaches a maximum, $X^{\mu}_{\rm max}$, which is not consistent with observations. This can be observed in figure \ref{lnA} where $X_{\rm max}$ and $X^{\mu}_{\rm max}$ have been converted (according with the tested hadronic models) to the average logarithmic mass $\langle \ln A \rangle$, as it can be seen in Fig. \ref{lnA}.

We conclude that post-LHC models still need further improvements that might hide unaccounted phenomena. By exploring this physics, one could also improve the uncertainty of the modelling of $X_{\rm max}$ itself.

\section{Prospects:AugerPrime}

The Pierre Auger Observatory is embarked in an upgrade of the Observatory called AugerPrime, which consists of new fastest electronics, improved FD uptime, and specially installing scintillators on top of the Cherenkov tanks which would allow a better discrimination between electromagnetic and muonic component in the ground. The basic scientific goals are:
1) Understand the origin of the flux suppression, whether it is caused by energy losses of cosmic rays or energy exhaustion of the sources. 2) Charged Particle Astronomy, by separating heavy from light primaries, which are less deflected by EGMF. 3) Also, make fundamental physics at ultra-high energies, by constraining the phase-space parameters in the hadronic multi-particle production.

\section*{Acknowledgments}

L. Cazon wants to thank funding by Fundac\~ ao para
a Ci\^e ncia e Tecnolog\' \i a, COMPETE, QREN, and European
Social Fund.

\end{document}